\documentclass[preprint,prd,nofootinbib,tightenlines,amsmath]{revtex4}
\usepackage{enumerate}
\usepackage{multirow}

\usepackage{epsfig}
\usepackage{graphics,color}      % usual driver
\usepackage{verbatim}
\usepackage{amsmath}    % need for subequations
\catcode`\@=11
\def\sla@#1#2#3#4#5{{%
 \setbox\z@\hbox{$\m@th#4#5$}%
 \setbox\tw@\hbox{$\m@th#4#1$}%
 \dimen4\wd\ifdim\wd\z@<\wd\tw@\tw@\else\z@\fi
 \dimen@\ht\tw@
 \advance\dimen@-\dp\tw@ \advance\dimen@-\ht\z@
 \advance\dimen@\dp\z@
 \divide\dimen@\tw@ \advance\dimen@-#3\ht\tw@
 \advance\dimen@-#3\dp\tw@ \dimen@ii#2\wd\z@
 \raise-\dimen@\hbox to\dimen4{%
 \hss\kern\dimen@ii\box\tw@\kern-\dimen@ii\hss}%
 \llap{\hbox to\dimen4{\hss\box\z@\hss}}}}

\def\slashed#1{%
 \expandafter\ifx\csname sla@\string#1\endcsname\relax
{\mathpalette{\sla@/00}{#1}}
\fi}
\def\declareslashed#1#2#3#4#5{%
 \expandafter\def\csname sla@\string#5\endcsname{%
#1{\mathpalette{\sla@{#2}{#3}{#4}}{#5}}}}
 \catcode`\@=12
\declareslashed{}{/}{.08}{0}{D}
 \declareslashed{}{/}{.1}{0}{A}
 \declareslashed{}{/}{0}{-.05}{k}
 \declareslashed{}{/}{.1}{0}{\partial}
 \declareslashed{}{\not}{-.6}{0}{f}

\def\lsim{\mathrel {\vcenter {\baselineskip 0pt \kern 0pt
    \hbox{$<$} \kern 0pt \hbox{$\sim$} }}}
\def\gsim{\mathrel {\vcenter {\baselineskip 0pt \kern 0pt
    \hbox{$>$} \kern 0pt \hbox{$\sim$} }}}

\newcommand{\bea}{\begin{eqnarray}}
\newcommand{\eea}{\end{eqnarray}}

\begin{document}

\baselineskip=15pt
\preprint{}

\title{Probing lepton gluonic couplings at the LHC}

\author{Hugh Potter and German Valencia}

\email{valencia@iastate.edu}

\affiliation{Department of Physics, Iowa State University, Ames, IA 50011.}

\date{\today}

\vskip 1cm
\begin{abstract}

In an effective Lagrangian description of physics beyond the standard model, gluonic couplings of leptons that respect the symmetries of the standard model occur in operators of dimension eight. Normally the effect of such operators is much suppressed in processes that occur at energies below the scale of new physics. For this reason usual studies of the effective Lagrangian are limited to the lowest dimension operators which is typically six. We point out that the large parton luminosity for gluon-gluon interactions at the LHC places these gluonic couplings to leptons within observable reach for a new physics scale near one TeV.

\end{abstract}

\pacs{PACS numbers: }

\maketitle

A general framework to study physics beyond the standard model (SM) is provided by the effective Lagrangian. Within this formalism one assumes that there is some new physics that appears at a high energy scale $\Lambda$ that can be integrated out and represented by an effective Lagrangian valid for energies below $\Lambda$. The dominant effects of the new physics are described by the lowest dimension operators. A complete catalog of effective operators up to dimension six that respect the symmetries of the SM has been presented in Ref.~\cite{Buchmuller:1985jz}, and many of them have been used for collider phenomenology. 

For example, the CDF collaboration studied the production of lepton pairs and by constraining its deviation from the Drell-Yan cross-section was able to place bounds on a dimension six operator related to electron-quark compositeness \cite{Abe:1991nd}. The same study was also used to constrain $Z^\prime$ bosons, illustrating the complementarity of direct searches for new resonances and indirect searches for their low energy (below the resonance) effects using the effective Lagrangian. More recently we have proposed to look for the tau-lepton charge asymmetry at the LHC as a way to distinguish different scenarios that can give rise to similar effects in the lepton pair invariant mass distribution \cite{Gupta:2011vt}.

The effects of dimension six operators in processes with an energy E below the new physics scale $\Lambda$ are suppressed by $E^2/\Lambda^2$. In the same manner higher dimension operators are further suppressed by additional powers of $E/\Lambda$ and for this reason most studies are limited to the lowest dimension operators. At the LHC, however, there is a very large parton luminosity for gluon-gluon interactions \cite{Eichten:1984eu,Quigg:2009gg} enhancing processes initiated by gluon fusion with respect to those initiated by $q\bar{q}$ annihilation. For certain parameter ranges, this enhancement in sufficient to alter the simple power counting. In this paper we address the question of the LHC sensitivity to gluonic couplings of leptons, which first appear in operators of dimension eight in the effective Lagrangian.

Any effective coupling between leptons and gluons must be in a color singlet configuration and we can use the ingredients of Ref.~\cite{Buchmuller:1985jz} to construct it. The leading operators in the effective Lagrangian appear at dimension eight and are of the form
\begin{eqnarray}
{\cal L} = \frac{g_s^2}{\Lambda^4}\left(c \ G^{A\ \mu\nu}G^A_{\mu\nu} \bar \ell_L \ell_R \phi  +\tilde{c} \ G^{A\ \mu\nu} \tilde G^A_{\mu\nu} \bar \ell_L \ell_R \phi \right)\ + {\rm h.~c.}
\end{eqnarray}
where $G^A_{\mu\nu}$ is the gluon field strength tensor and $\tilde G^{A\mu\nu} = (1/2)\epsilon^{\mu\nu\alpha\beta}G^A_{\alpha\beta}$.
If we allow for CP violating phases in the coefficients, $ce^{i\phi}$ and $\tilde{c}e^{i\tilde{\phi}}$, the resulting gluon-lepton couplings not involving a Higgs boson are 
\begin{eqnarray}
{\cal L} &=& \frac{v g_s^2}{\Lambda^4}\left(c\ \cos\phi \ G^{A\ \mu\nu}G^A_{\mu\nu}  +\tilde{c}\ \cos\tilde{\phi} \ G^{A\ \mu\nu} \tilde G^A_{\mu\nu}\right) \bar \ell \ell \nonumber \\
&+&   \frac{iv g_s^2}{\Lambda^4}\left(c\ \sin\phi \ G^{A\ \mu\nu}G^A_{\mu\nu}  +\tilde{c}\ \sin\tilde{\phi} \ G^{A\ \mu\nu} \tilde G^A_{\mu\nu}\right) \bar \ell \gamma_5 \ell
\label{couplings}
\end{eqnarray}
where $v$ is the Higgs vacuum expectation value $v\sim 246$~GeV. The couplings $c,\ \tilde{c}$ (as well as the leptons $\ell$) in this equation are understood to carry a generation index, and are in principle different for electrons, muons or tau-leptons. They could be lepton number violating as well.

\begin{figure}[th]
\centerline{
\includegraphics[angle=270,width=.8\textwidth]{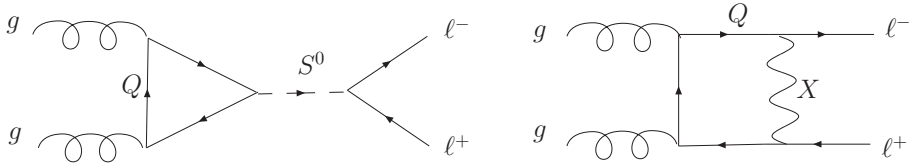}
} \caption{Sample Feynman diagrams giving rise to the operators in Eq.~\ref{couplings} at the one-loop level as described in the text. }\label{fig:diagrams}
\end{figure}

There are several models that would produce these couplings. In Figure~\ref{fig:diagrams} we illustrate two generic cases where this happens at one-loop.
The figure on the left already occurs in the SM where it corresponds to gluon fusion production of a Higgs boson followed by its decay into leptons. In fact  this is a  process already used by the ATLAS \cite{Aad:2011rv} and CMS \cite{Chatrchyan:2011nx} collaborations to search for MSSM Higgs bosons in the $\tau^+\tau^-$ channel. For the scenario that we have in mind, however, the $S^0$ is a heavy neutral scalar (with mass in the TeV range) and its couplings to leptons would have to be large to generate a sizable effective coupling. To sketch an example we could think of a 2HDM with large $\tan\beta$ in a scenario with four generations of quarks. The heavy quark in the loop $Q$ could be a $b^\prime$ and this diagram would result in effective couplings of order $c \sim m_\ell \tan^2\beta/v$ with $\Lambda^4 \sim (4\pi v)^2 M_S^2$. The figure on the right illustrates another possibility; $Q$ is again a heavy quark and $X$ a lepto-quark with appropriate quantum numbers \cite{leptoquarks}. For a strongly coupled vector lepto-quark this diagram would result in effective couplings of order $c \sim \pi \alpha_s$ with $\Lambda^4 \sim (4\pi v)^2 M_X^2$. A generic feature of these couplings when induced at one-loop is that they involve in principle two different heavy scales: $4\pi v$ and the mass of the new particle. We are not aware of any existing low energy bounds on these couplings from their contributions to rare meson decays. It would be straightforward to map the operators of Eq.~\ref{couplings} into the low energy chiral Lagrangian using the same   techniques appropriate for couplings of light Higgs bosons \cite{efflag}. However, simple dimensional analysis tells us that these operators will induce corrections to  low energy amplitudes of about the same size as second order weak amplitudes which are  below the sensitivity that can be attained with rare meson or hyperon decays.

Turning our attention to high energy colliders, we use the couplings in Eq.~\ref{couplings} to compute the differential and total cross-section at the parton level. Neglecting the lepton mass we find 
\begin{eqnarray}
\frac{d\hat{\sigma}(gg\to \ell^+\ell^-)}{d\hat{t}}&=&\left(|c|^2+|\tilde{c}|^2\right)\frac{v^2 g_s^4}{\Lambda^8}\frac{\hat{s}}{32\pi}, \nonumber \\
\hat{\sigma}(gg\to \ell^+\ell^-)&=&\left(|c|^2+|\tilde{c}|^2\right)\frac{v^2 g_s^4}{\Lambda^8}\frac{\hat{s}^2}{32\pi}.
\label{sigmagg}
\end{eqnarray}
This result indicates that the two types of coupling $c,\ \tilde{c}$ have the same effect and cannot be distinguished by studying the lepton pair production process alone. The situation is familiar from Higgs studies, so we know that it is also possible in principle to examine the $CP$ violating interference between the $c$ and $\tilde{c}$ couplings for the case of tau-lepton pairs if we also study the subsequent tau-lepton decay \cite{cptau}. We do not consider this possibility  and will restrict ourselves to the study of $c$ for the remainder of this paper. 

It is useful to compare the cross-section of Eq.~\ref{sigmagg} to one that is due to new physics that appears in a dimension six operator and contributes to lepton pair production in a light $q\bar{q}$ annihilation initiated process. There are many such operators, but for our current purpose it will suffice to  choose the operator (contained in ${\cal O}_{\ell q}$ with form $\bar{L}R\bar{L}R$ of Ref.~\cite{Buchmuller:1985jz})\footnote{Similar four-fermion operators connecting quarks and neutrinos have been considered recently in the context of monojets at the LHC \cite{Friedland:2011za}.}
\begin{eqnarray}
{\cal L}=a \ \frac{g^2}{\Lambda^2}\ \bar{u}u\ \bar{\ell}\ell.
\label{qqcoups}
\end{eqnarray}
The Dirac structure of this operator ensures that it does not interfere with the dominant ($\gamma$ and $Z$ mediated) SM amplitudes, and that it has the same angular distribution as Eq.~\ref{sigmagg}, facilitating our comparisons. The resulting parton level differential and total cross-sections are
\begin{eqnarray}
\frac{d\hat\sigma(u\bar{u}\to \ell^+\ell^-)}{d\hat{t}}=\frac{|a|^2 g^4 }{48\pi\Lambda^4},
&&
\hat\sigma(u\bar{u}\to \ell^+\ell^-)=\frac{|a|^2 g^4 \hat{s}}{48\pi\Lambda^4}.
\label{qqsigma}
\end{eqnarray}
We first comment on the energy dependence of the results in Eq.~\ref{sigmagg}~and~Eq.~\ref{qqsigma} which is unlike the behavior of the QED (Drell-Yan) cross-section
\begin{eqnarray}
\hat{\sigma}(q\bar{q}\to \ell^+\ell^-)=\frac{Q_q^2}{3}\frac{4\pi\alpha^2}{3\hat{s}}.
\end{eqnarray}
The growth with $\hat{s}$, linear for Eq.~\ref{qqsigma} and quadratic for Eq.~\ref{sigmagg}, exhibited by these results reveals the nature of the effective theory. These results are only valid for energies {\it below} the scale of new physics $\Lambda$. This means that relative to the Drell-Yan cross-section,  Eq.~\ref{qqsigma} is {\it suppressed} by $\hat{s}^2/\Lambda^4$ and Eq.~\ref{sigmagg}  is {\it suppressed} by an additional $v^2\hat{s}/\Lambda^4$. The latter would make effects due to dimension eight operators much smaller than those due to dimension six operators in accordance with the power counting. Specifically, for the two processes under discussion,
\begin{eqnarray}
[\hat{s}\hat{\sigma}(u\bar{u}\to \ell^+\ell^-)]_{Eq.~\ref{qqsigma}} \sim  \left(\frac{\hat{s}}{\Lambda^2}\right)^2, &&
[\hat{s}\hat{\sigma}(gg\to \ell^+\ell^-)]_{Eq.~\ref{sigmagg}} \sim  \left(\frac{\hat{s}}{\Lambda^2}\right)^3 
\end{eqnarray}
with proportionality constants that are of the same order for $a\sim c$. For an  LHC analysis using the energy range $150 \lsim \sqrt{\hat{s}} \lsim 300$~GeV and assuming new physics at one TeV, one can see that the relative suppression in $\hat{\sigma}(gg\to \ell^+\ell^-)$ will be roughly countered by the relative enhancement of  its parton luminosity \cite{Quigg:2009gg}. 

To make the previous statement more precise we now present our numerical results.
To this end we implement the effective couplings of Eq.~\ref{couplings}~and~Eq.~\ref{qqcoups} into  {\tt MadGraph 5} \cite{madgraph} using FeynRules \cite{Christensen:2008py}. We use  {\tt MadGraph 5} \cite{madgraph} for our event generation with {\tt CTEQ6L-1} parton distribution functions~\cite{Pumplin:2002vw}, and we apply  the basic acceptance cuts: $p_{T_\ell} > 20$ GeV, $\left|\eta_\ell \right| < 2.5$, and $\Delta R_{\ell\ell} > 0.4$. We  also assume the LHC is running at $\sqrt{S} = 14$~TeV  center-of-mass energy, with an integrated luminosity of $10~{\rm fb}^{-1}$ per year to estimate the sensitivity.\footnote{ We have also performed some elementary  checks on our use of FeynRules and {\tt MadGraph 5}   by comparing with a direct implementation of the vertices in Comphep \cite{comphep}. } 

\begin{figure}[th]
\centerline{
\includegraphics[angle=-90,width=.55\textwidth]{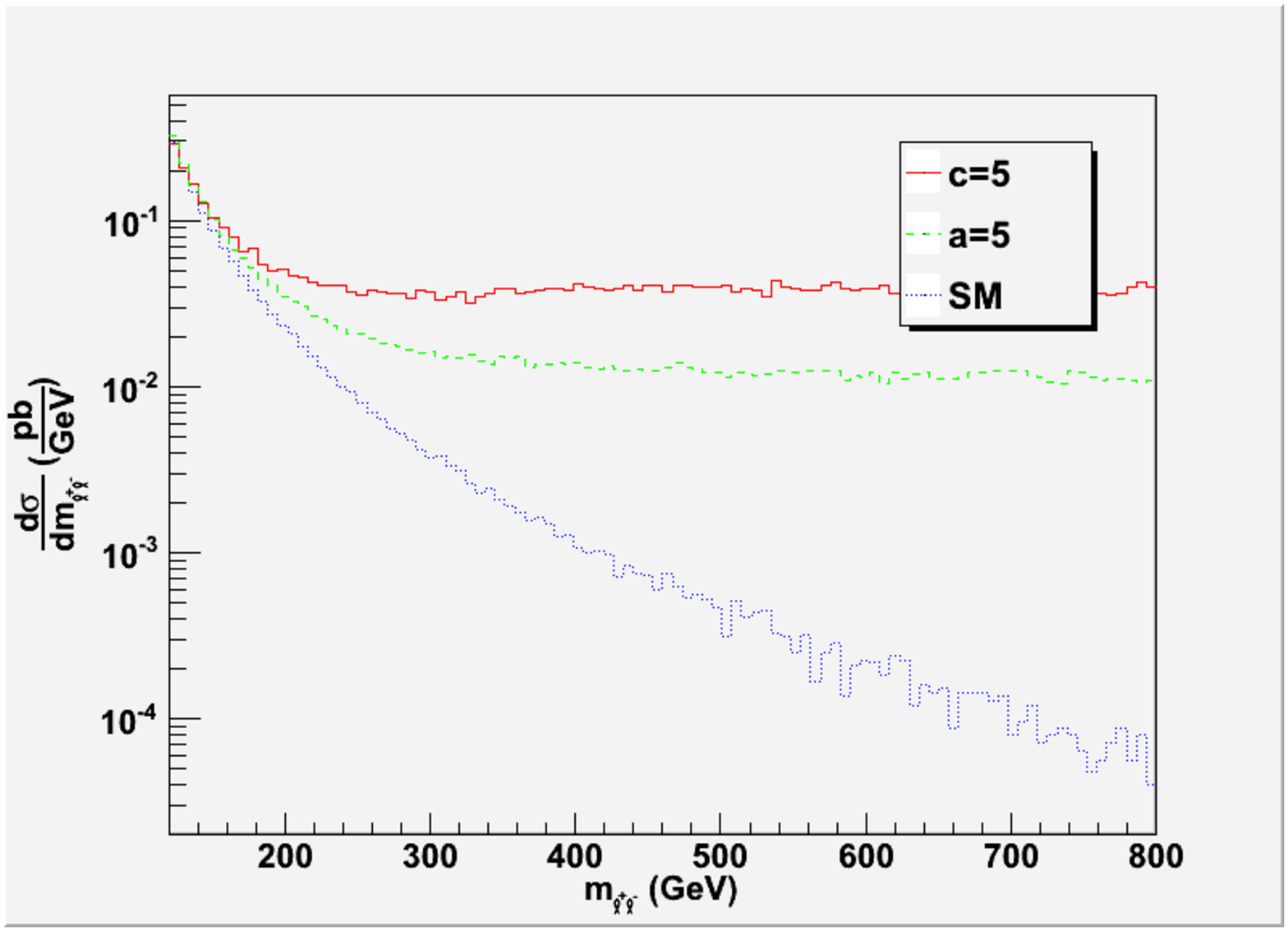}\hspace{.3in}
\includegraphics[angle=-90,width=.55\textwidth]{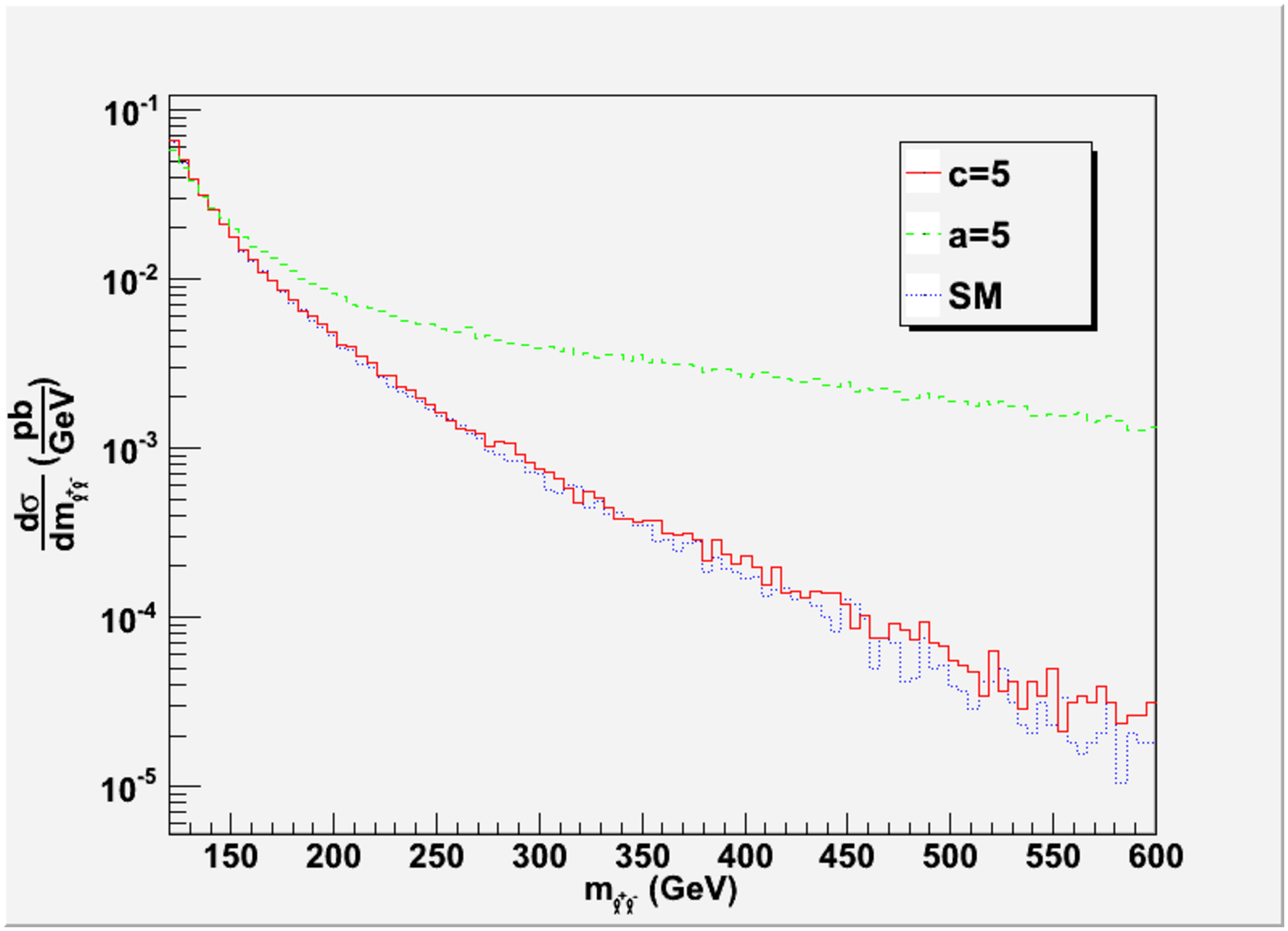}
}
\caption{$d\sigma/dm_{\ell\ell}$ for the SM; the SM plus new physics in the gluon fusion process ($c=5$); and the SM plus new physics in the $u\bar{u}$ annihilation process ($a=5$). The scale of new physics is taken to be $\Lambda =1$~TeV. The figure on the left corresponds to $pp\to \ell^+\ell^-$ at the LHC and the figure at the right to $p\bar{p}\to \ell^+\ell^-$ at the Tevatron. }
\label{fig:mll}
\end{figure}

In Figure~\ref{fig:mll} we illustrate the invariant mass distributions $d\sigma/dm_{\ell\ell}$ for  $pp\to \ell^+\ell^-$ at the LHC and  $p\bar{p}\to \ell^+\ell^-$ at the Tevatron. The blue (dotted) curve corresponds to the SM; the red (solid) curve adds the dimension eight operator of Eq.~\ref{couplings} with $c=5$ and $\Lambda =1$~TeV to the SM; and the green (dashed) curve adds the dimension six operator of Eq.~\ref{qqcoups} with $a=5$ and $\Lambda=1$~TeV to the SM.
For the parameters chosen, Figure~\ref{fig:mll} indicates that the gluon fusion initiated process results in a larger correction to the SM  than the $u\bar{u}$ initiated process at the LHC even though the former originates in a higher dimension operator. The reverse is true for the Tevatron, where the dimension eight correction to gluon fusion is negligible. 

To estimate the sensitivity to these new couplings we need to compare the number of events to those expected in the SM. It is therefore advantageous to first remove most of the SM events with a cut $m_{\ell\ell}\geq 120$~GeV and to concentrate on the region above the $Z$ peak. Recalling that the effective theory is only valid below the scale $\Lambda$, we consider two cases:
\begin{itemize}

\item A relatively low new physics scale $\Lambda \sim 1$~TeV. In this case we limit the comparison to the range $120 \leq m_{\ell\ell}\leq 300$~GeV. The cross-section for these cuts is approximately
\begin{eqnarray}
\sigma(pp\to \ell^+\ell^-) \approx \left(8.5 + 0.17\ c^2 \left(\frac{1~{\rm TeV}}{\Lambda}\right)^8+ 0.077 \ a^2  \left(\frac{1~{\rm TeV}}{\Lambda}\right)^4\right) pb
\end{eqnarray}

The statistical sensitivity to the signal in this region is simply given by the number of signal events (new physics) over the square root of the number of background events (SM):
\begin{eqnarray}
S=\frac{\sigma_S}{\sqrt{\sigma_{SM}}}\sqrt{{\cal L}}
\end{eqnarray}
where ${\cal L}$ is the total integrated luminosity. For illustration we require $3\sigma$ statistical sensitivity for one year of nominal LHC running, 10~fb$^{-1}$. With this we find that the LHC is sensitive to 
\begin{eqnarray}
c \gsim 0.7  , &&  a \gsim 1.1 .
\end{eqnarray}

\item For a somewhat larger scale $\Lambda \sim 2$~TeV we can use the range $120 \leq m_{\ell\ell}\leq 1000$~GeV  to estimate the sensitivity. 
The cross-section for these cuts is approximately
\begin{eqnarray}
\sigma(pp\to \ell^+\ell^-) \approx \left(9.0 + 1.23\ c^2 \left(\frac{1~{\rm TeV}}{\Lambda}\right)^8+ 0.4 \ a^2  \left(\frac{1~{\rm TeV}}{\Lambda}\right)^4\right) pb
\label{lhcnum}
\end{eqnarray}
With the same criteria as above we find
\begin{eqnarray}
c \gsim 4.3  , &&  a \gsim 1.9.
\end{eqnarray}

\end{itemize}
The fact that the coupling $c$ corresponds to a dimension eight operator is responsible for the faster drop in sensitivity, compared to $a$, when the scale of new physics increases.

We could repeat the analysis of Eq.~\ref{lhcnum} for the Tevatron, even though Figure~\ref{fig:mll} already suggests that the bounds will be significantly weaker. With the same parameters used in Eq.~\ref{lhcnum} we find that for $\Lambda=2$~TeV, the Tevatron can place the constraint $|c| \lsim 75$. 

Similarly we can use the LEP-II measurements of $\sigma(e^+e^-\to {\rm hadrons})$ to constrain this interaction for the case of gluonic couplings of electrons. The LEP measurements agree very well with the SM prediction, so we obtain a bound by requiring that the new contribution be less than three times the experimental error. Since the new contribution increases with center of mass energy,
\begin{equation}
\sigma(e^+e^- \to gg) = 32 \pi \alpha_s^2 \left(|c|^2+|\tilde{c}|^2\right)\frac{v^2s^2}{\Lambda^8},
\end{equation}
we extract a bound using the measurement at $\sqrt{s}=202$~GeV which is near the upper reach of LEP and where the experimental error is somewhat larger. From Ref.~\cite{Abbaneo:2001ix}, for $\sqrt{s}=202$~GeV we find 
\begin{eqnarray}
\sigma(q\bar{q})=(19.278\pm0.430){\rm ~pb} && \sigma(q\bar{q})_{SM}=18.572 {\rm ~pb}\nonumber \\
\sigma(e^+e^-)\to gg \leq 1.29 {\rm~ pb} &\implies& |c| \lsim 80.
\end{eqnarray}
Using the other values of $\sqrt{s}$ measured by LEP-II results in very similar numbers.

Finally we can compute the zero'th partial wave for $gg \to \ell^+\ell^-$ to obtain a unitarity limit by requiring ${\rm Re~}{a_0}<1/2$,
\begin{equation}
a_0=\frac{c}{2}\alpha_s\frac{vs^{3/2}}{\Lambda^4}\leq\frac{1}{2} \implies |c| \lsim 80
\end{equation}
where the numerical bound follows from requiring that the partial wave satisfy the unitarity constraint up to $\sqrt{s} = \Lambda=2$~TeV.

In conclusion we have shown that the LHC is in principle sensitive to gluonic couplings of leptons provided these are induced by new physics at a scale not far above 1~TeV.  To our knowledge, these couplings have not been significantly  constrained by low energy processes and the LHC offers the first opportunity to study them. Of course, to extract these bounds it will be necessary to compare the measured cross-sections to sufficiently precise SM calculations. Alternatively one can use only experimental data to test the lepton-flavor dependence of these couplings by comparing electron and muon pair production, for example. Experimentally, one should also test the possibility of lepton flavor violation in these new couplings.

\begin{acknowledgments}

This work was supported in part by the DOE under contract number
DE-FG02-01ER41155. We thank Sudhir Kumar Gupta and Natascia Vignaroli for their help with FeynRules and Madgraph, and David Atwood for comments on the manuscript.

\end{acknowledgments}

\end{document}